\documentclass[aps,physrev,reprint,superscriptaddress,raggedbottom]{revtex4-2}

\usepackage{braket}
\usepackage{amsmath}
\usepackage{graphicx}
\usepackage[colorlinks=true, linkcolor=blue, citecolor=blue, urlcolor=blue]{hyperref}

\begin{document}

\title{XENONnT \texorpdfstring{$pp$}{pp}-Dominated Solar Neutrinos Constrain the \texorpdfstring{$\nu_1$}{nu1} Pseudo-Dirac Splitting}

\author{Gio Leone}
\email[]{grleone@uchicago.edu}
\affiliation{Department of Physics, Kavli Institute for Cosmological Physics, and Enrico Fermi Institute; University of Chicago, Chicago, IL 60637, USA}
\date{\today}

\begin{abstract}
XENONnT has measured a low-energy solar-neutrino signal dominated by $pp$ neutrinos. At these energies, an ultra-small pseudo-Dirac mass-squared splitting induces detectable active-sterile oscillations over the Sun--Earth baseline. We study pseudo-Dirac oscillations of the first mass eigenstate, compute the resulting modification of the $pp+{}^7$Be electron-recoil spectrum, and construct a likelihood from the published XENONnT energy spectra. We find that the XENONnT data exclude $5.9\times10^{-13}\,\mathrm{eV}^2\lesssim \delta m_1^2 \leq 10^{-11}\,\mathrm{eV}^2$ at the $2\sigma$ level, extending current solar-neutrino sensitivity to smaller mass splittings and approaching a regime previously discussed for future $pp$-sensitive detectors.
\end{abstract}

\maketitle

\section{Introduction}

Solar neutrinos are precision probes of flavor conversion and propagation~\cite{Haxton:2012wfz}. The Sun prepares an energy-dependent mixture of neutrino mass eigenstates, which subsequently propagates over the Sun--Earth baseline~\cite{Wolfenstein:1977ue,Mikheyev:1985zog,Parke:1986jy}. As they travel, solar neutrinos can oscillate between different combinations of mass eigenstates. The particular mixture of sub-MeV energies and the approximately $1\,$AU baseline gives $pp$ and ${}^7$Be neutrinos sensitivity to oscillation phases generated by extremely small mass-squared splittings.

A pseudo-Dirac neutrino is an approximate Dirac state split into two nearly degenerate Majorana mass eigenstates by a small lepton-number-violating Majorana mass~\cite{Wolfenstein:1981kw,Petcov:1982ya,Kobayashi:2000md}. In the pseudo-Dirac limit, where the Majorana mass is much smaller than the Dirac mass, the two components are maximally mixed combinations of active and sterile neutrinos. Their small mass-squared splitting thus generates long-baseline active--sterile neutrino oscillations that reduce the active neutrino flux~\cite{deGouvea:2009fp,Anamiati:2017rxw}. Pseudo-Dirac effects have been studied using terrestrial oscillation experiments, solar neutrinos, supernova neutrinos, and high-energy astrophysical neutrinos~\cite{Beacom:2003eu,Esmaili:2009fk,deGouvea:2009fp, Anamiati:2017rxw, Anamiati:2019maf,Martinez-Soler:2021unz,Ansarifard:2022kvy,Carloni:2022cqz,Rink:2022nvw,Carloni:2025dhv}.

The recent XENONnT measurement provides a solar-neutrino electron-recoil sample that is approximately $89\%$ $pp$ and $10\%$ ${}^7$Be in the $1$--$140\,\mathrm{keV}$ reconstructed-energy region~\cite{XENON:2026rbz}. The combination of a $pp$-dominated spectrum and a measured absolute solar-neutrino normalization makes this data set particularly sensitive to pseudo-Dirac oscillations. We focus on a pseudo-Dirac splitting of the first mass eigenstate, denoted $\delta m_1^2$. An analysis by Ansarifard and Farzan combining Borexino low-energy solar-neutrino rates with Super-Kamiokande ${}^8$B spectral data obtains a constraint reaching $\delta m_1^2 \simeq 1.5\times10^{-12}\,\mathrm{eV}^2$ at $2\sigma$~\cite{Ansarifard:2022kvy,Borexino:2017rsf,Super-Kamiokande:2016yck}. Future solar-neutrino measurements are expected to extend sensitivity to smaller mass splittings. de Gouv\^ea et al.\ forecast future DARWIN sensitivity to $\delta m_1^2$ at the $\mathcal O(10^{-13}\,\mathrm{eV}^2)$ level~\cite{deGouvea:2021ymm, DARWIN:2020bnc}. Franklin, Perez-Gonzalez and Turner similarly forecast sensitivity to $\delta m_1^2$ at the $\mathcal O(10^{-13}\,\mathrm{eV}^2)$ level if JUNO can access the high-energy portion of the $pp$ spectrum~\cite{Franklin:2023diy,JUNO:2023zty}.

In this work, we present, to the best of our knowledge, the first constraint on $\delta m_1^2$ using an observed $pp$-dominated solar-neutrino sample in a dark-matter detector. The splitting $\delta m_1^2$ modifies the active-flavor probabilities at Earth, deforming the solar neutrino electron-recoil spectra. We perform a likelihood analysis on digitized versions of the published XENONnT SR0 and SR1b spectra~\cite{XENON:2026rbz}. After including reconstructed background components and profiling the associated nuisance parameters, we find that the best fit occurs at zero pseudo-Dirac splitting and exclude $5.9\times10^{-13}\,\mathrm{eV}^2\lesssim \delta m_1^2 \leq 10^{-11}\,\mathrm{eV}^2$ at the $2\sigma$ level.

\section{Pseudo-Dirac solar neutrinos}
\subsection{Minimal Model}

We consider four vacuum mass eigenstates $\ket{\nu_i}$ ($i=1,2,3,4$), with $\ket{\nu_1}$ and $\ket{\nu_4}$ forming a nearly degenerate pseudo-Dirac pair. In the pseudo-Dirac limit, the dominant Dirac mass is perturbed by a much smaller Majorana mass. The two nearly-degenerate eigenstates $\ket{\nu_1}$ and $\ket{\nu_4}$ have masses we denote as $m_1$ and $m_4$, respectively, where we take $m_4> m_1$. Let $\delta m_1^2 \equiv m_4^2-m_1^2$ denote their mass-squared splitting. Up to phase conventions, the active and sterile combinations associated with this pair, labeled $\ket{\nu^a_1}$ and $\ket{\nu^s_1}$ respectively, can be expressed in terms of the vacuum mass eigenstates as
\begin{align}
    \label{eq:states}
    &\ket{\nu^a_1} \simeq \frac{\ket{\nu_1}+\ket{\nu_4}}{\sqrt 2}, &
    &\ket{\nu^s_1} \simeq \frac{\ket{\nu_1}-\ket{\nu_4}}{\sqrt 2}.&
\end{align}
Thus, the active and sterile combinations are approximately the symmetric and antisymmetric combinations of the two nearly-degenerate mass eigenstates.

\begin{figure}[t]
    \centering
    \includegraphics[width=.9\linewidth]{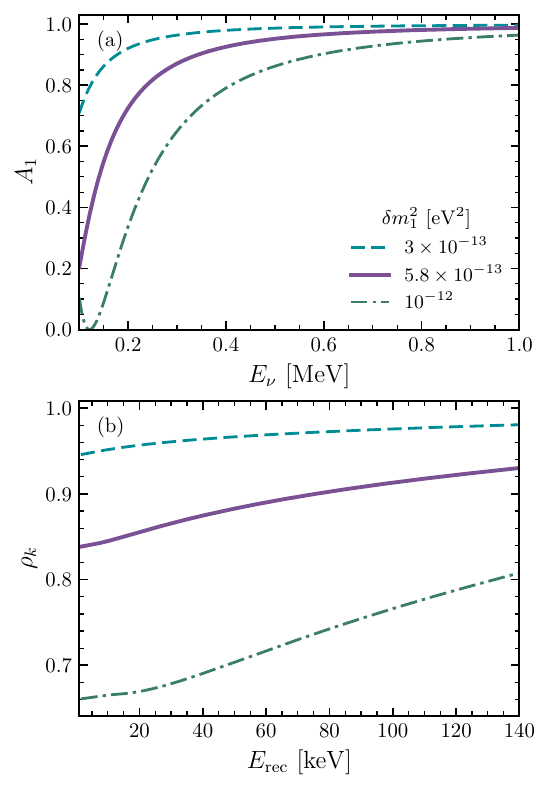}
    \caption{(a) The active-survival factor of the first pseudo-Dirac pair for various pseudo-Dirac mass-squared splittings $\delta m_1^2$. We take the baseline $L = 1\,$AU. The sub-MeV energy range remains sensitive to the pseudo-Dirac oscillation phase. (b) Expected deformation of the reconstructed solar-neutrino spectrum for the same mass-squared splittings.}
    \label{fig:pd_deformation}
\end{figure}

After a baseline propagation of length $L$, the two mass components $\ket{\nu_1}$ and $\ket{\nu_4}$ acquire phases $\exp[-i m_{1,4}^2L/(2E_{\nu})]$~\cite{Giunti:1992hk}. $E_{\nu}$ is the common ultra-relativistic energy of the propagating neutrino wavepacket, which is well approximated by the common energy of the nearly-degenerate Majorana components in the pseudo-Dirac limit. The probability that the active component $\ket{\nu^a_1}$ remains active over the baseline $L$ is
\begin{equation}
\label{eq:active survival}
    A_1(E_\nu,L) = \cos^2\!\left(\frac{\delta m_1^2 L}{4 E_\nu}\right).
\end{equation}
In the minimal model, the active-survival factor of the first pseudo-Dirac pair $A_1$ is nontrivial (see Fig.~\ref{fig:pd_deformation}), whereas the active-survival factors associated with the other two standard mass components are $A_2=A_3=1$.

For a small pseudo-Dirac oscillation phase, the probability that the active component oscillates into its sterile partner follows
\begin{equation}
    \label{eq:sterileProbability}
    1-A_1(E_\nu,L) \simeq \left(\frac{\delta m_1^2 L}{4E_\nu}\right)^2,
\end{equation}
and is hence proportional to $L^2/E_\nu^2$ for fixed $\delta m_1^2$. Therefore, low-energy neutrinos are sensitive to sterile oscillations over large baselines even for very small pseudo-Dirac mass-squared splittings. Sub-MeV solar neutrinos are particularly sensitive to such splittings over the Sun--Earth baseline.

The Earth--Sun distance varies annually by about $1.7\%$ about its mean value of $1\,$AU, which in principle modulates the pseudo-Dirac oscillation phase~\cite{BOREXINO:2022wuy}. Since the oscillation phase is proportional to $L$, its fractional variation follows the orbital variation. XENONnT does not publish the time-resolved exposure needed to construct the corresponding exposure-weighted baseline for our recast. We thus take $L=1\,$AU throughout our analysis and neglect orbital averaging. We therefore conservatively restrict our inference to $\delta m_1^2 \leq 10^{-11}\,\mathrm{eV}^2$, beyond which Sun--Earth baseline averaging become increasingly important and motivates a dedicated time-dependent treatment \cite{Ansarifard:2022kvy}.

\subsection{Solar Propagation}

Solar weak reactions produce electron-flavor neutrinos~\cite{Haxton:2012wfz,Vinyoles:2016djt}. 
We take the central normal-ordering oscillation parameters from NuFIT~\cite{Esteban:2024eli} and let $U$ denote the vacuum three-neutrino PMNS matrix. At a production point $r$ inside the Sun, the electron-flavor state can be expanded in the instantaneous matter eigenbasis,
\begin{equation}
    \ket{\nu_e} = \sum_{i=1}^3 U^{m*}_{ei}(E_\nu,r) \ket{\nu_i^m(E_\nu,r)},
\end{equation}
where $U^m$ is the matter mixing matrix and $\ket{\nu_i^m(E_\nu,r)}$ denotes the $i$-th instantaneous matter eigenstate.

In the adiabatic LMA-MSW regime relevant here, each instantaneous matter eigenstate evolves continuously as the solar density decreases and, in the standard three-flavor limit, approaches the corresponding vacuum mass eigenstate $\ket{\nu_i}$ near the solar surface~\cite{Wolfenstein:1977ue,Mikheyev:1985zog,Parke:1986jy,Haxton:2012wfz}. After averaging over the extended production region and the rapidly varying standard oscillation phases, the three standard mass components are treated incoherently~\cite{Giunti:1992hk,deGouvea:2009fp,Ansarifard:2022kvy}. 
For a solar source $X$ with a normalized radial production profile $f_X(r)$, the source-averaged weight of the $i$-th standard three-flavor mass component is
\begin{equation}
    \label{eq:neutrinoEmergence}
    p^{X,\odot}_i(E_\nu) = \int_0^{R_\odot} dr\, f_X(r) |U^m_{ei}(E_\nu,r)|^2,
\end{equation}
where $R_\odot = 6.96\times10^5~\mathrm{km}$ is the solar radius and $\sum_{i=1}^3 p^{X,\odot}_i(E_\nu) = 1$~\cite{deGouvea:2021ymm,Ansarifard:2022kvy}. For the electron density entering the MSW potential, we use the analytic solar electron-density profile of Ref.~\cite{Bahcall:2005va}, as implemented in SNuDD~\cite{Amaral:2026gml}.

For the ultra-small pseudo-Dirac mass-squared splittings considered here, we neglect corrections from $\delta m_1^2$ to the standard LMA-MSW preparation of the three active mass components. We evaluate the solar production weights $p_i^{X,\odot}$ using the standard low-energy three-flavor LMA-MSW approximation, treating the $1$--$2$ matter mixing explicitly while retaining the vacuum value of $\theta_{13}$ and associating the first standard mass component with the active combination $\ket{\nu_1^a}$. The relative phase within the nearly degenerate $\nu_1$--$\nu_4$ pair is then accumulated during propagation from the Sun to Earth and described by the active-survival factor $A_1(E_\nu,L)$ in Eq.~\eqref{eq:active survival}~\cite{deGouvea:2009fp}.

For $pp$ neutrinos, $E_\nu \leq 0.42\,\mathrm{MeV}$, and standard solar matter effects are small~\cite{deGouvea:2021ymm,Franklin:2023diy}. Nevertheless, for consistency we retain the matter correction and average $|U^m_{ei}(E_\nu,r)|^2$ over the $pp$ radial production distribution at each neutrino energy. We use the source-specific radial production distributions packaged with SNuDD~\cite{Amaral:2026gml, Lopes:2013nfa}. There are two ${}^7\mathrm{Be}$ lines, a dominant $861$-keV line and a subdominant $384$-keV line~\cite{Adelberger:2010qa}. At the dominant line, standard active-sector solar matter corrections are no longer negligible~\cite{Ansarifard:2022kvy,Franklin:2023diy}. We therefore treat both ${}^7\mathrm{Be}$ lines by averaging the matter-eigenstate production probabilities over the corresponding ${}^7\mathrm{Be}$ radial production distribution. The solar-neutrino flux normalizations are taken from the B16-GS98 standard solar model~\cite{Vinyoles:2016djt}. For a solar source $X$, these production probabilities enter the electron-flavor probability at Earth as
\begin{equation}
    \label{eq:prob_e}
    P_{e}^X(E_\nu)=\sum_{i=1}^3 p_i^{X,\odot}(E_\nu)|U_{ei}|^2A_i(E_\nu,L),
\end{equation}
while the combined probability to arrive as a non-electron active flavor, $P_{\mu\tau}^X\equiv P_{e\mu}^X+P_{e\tau}^X$, is~\cite{deGouvea:2021ymm,Ansarifard:2022kvy}
\begin{equation}
    \label{eq:prob_not_e}
    P_{\mu\tau}^X(E_\nu)=\sum_{i=1}^3 p_i^{X,\odot}(E_\nu)\left(1-|U_{ei}|^2\right)A_i(E_\nu,L).
\end{equation}

\section{XENON\texorpdfstring{\lowercase{n}}{n}T spectral recast}
\subsection{Solar-Neutrino Electron Recoils}

We obtain the $pp$ energy spectrum $d\Phi_{pp}/dE_{\nu}$ using the normalized Bahcall $pp$ spectral shape $\lambda_{pp}(E_\nu)$ and the $pp$ neutrino flux $\Phi_{pp} = 5.98\times 10^{10}\,\mathrm{cm}^{-2}\,\mathrm{s}^{-1}$ ~\cite{Bahcall:1997eg,Vinyoles:2016djt},
\begin{equation}
    \frac{d\Phi_{pp}}{dE_{\nu}} = \Phi_{pp}\lambda_{pp}(E_\nu).
\end{equation}
For ${}^7$Be, the spectrum is taken as the branching-fraction-weighted sum over the $E_L = 384.3\,\mathrm{keV}$ and $E_H=861.3\,\mathrm{keV}$ lines, approximating the energy spectrum of each monoenergetic branch as a delta function:
\begin{equation}
    \frac{d\Phi_{{}^7\mathrm{Be}}}{dE_\nu} = \Phi_{{}^7\mathrm{Be}}\left[b_L\,\delta(E_\nu-E_L) + b_H\,\delta(E_\nu-E_H) \right],
\end{equation}
where the branching fractions are $b_L \simeq 0.105$ and $b_H \simeq 0.895$~\cite{Adelberger:2010qa}. We take the ${}^7$Be neutrino flux to be $\Phi_{{}^7\mathrm{Be}} = 4.93\times 10^{9}\,\mathrm{cm}^{-2}\,\mathrm{s}^{-1}$~\cite{Vinyoles:2016djt}.

We calculate the $pp$ and ${}^7$Be electron-recoil spectra using the tree-level Standard Model neutrino--electron scattering cross sections~\cite{Baudis:2013qla}. Following XENONnT, we calculate the spectrum as a function of the true electron-recoil energy $E_r$ in the free-electron approximation (FEA)~\cite{XENON:2026rbz,Baudis:2013qla}. We construct the pseudo-Dirac recoil spectrum by replacing the standard flavor probabilities with the pseudo-Dirac probabilities in Eqs.~\eqref{eq:prob_e} and \eqref{eq:prob_not_e}. For a solar source $X$, the relevant recoil spectrum up to overall detector normalization is
\begin{equation}
    \label{eq:scattering_rate}
    \begin{aligned}
        \frac{dR_X}{dE_r} = \int dE_\nu\, \frac{d\Phi_X}{dE_\nu} \bigg[&P_e^X(E_\nu)\frac{d\sigma_e}{dE_r}+\\
            &\quad P_{\mu\tau}^X(E_\nu)\frac{d\sigma_{\mu\tau}}{dE_r}\bigg] Z_{\rm eff}(E_r).
    \end{aligned}
\end{equation}
Here $d\Phi_X/dE_\nu$ includes the total flux normalization and spectral shape of source $X$. The differential cross section $d\sigma_e/dE_r$ describes $\nu_e$--electron scattering, while $d\sigma_{\mu\tau}/dE_r$ denotes the common tree-level $\nu_\mu$--electron and $\nu_\tau$--electron differential cross section. The kinematically allowed neutrino-energy range for each recoil energy is understood in the $E_\nu$ integration.

We denote the total theoretical electron-recoil spectrum as
$dR/dE_r \equiv dR_{pp}/dE_r + dR_{{}^7{\rm Be}}/dE_r$.
We neglect the remaining solar components, which together contribute less than $1\%$ of the XENONnT solar-neutrino signal in the energy range of interest~\cite{XENON:2026rbz}. Following the primary XENONnT solar-neutrino signal model, we use the FEA with shell stepping to approximately account for xenon binding effects at low recoil energies~\cite{XENON:2026rbz,Baudis:2013qla}. Thus, the effective number of xenon electrons available for scattering at recoil energy $E_r$ is
\begin{equation}
    Z_{\rm eff}(E_r) = \sum_{s=1}^{Z} \Theta\!\left(E_r-E_b^{(s)}\right),
\end{equation}
where $Z=54$ is the atomic number of xenon, $\Theta$ is the Heaviside step function, and $E_b^{(s)}$ is the binding energy associated with the $s$-th xenon electron.

\subsection{Pseudo-Dirac Spectral Deformation}

We use the total theoretical electron-recoil spectrum $dR/dE_r$ to determine the relative spectral deformation induced by pseudo-Dirac oscillations. 
To compare the XENONnT fitted SM solar neutrino recoil spectrum with the spectrum expected for a pseudo-Dirac splitting $\delta m_1^2$, we define the binned theoretical recoil spectrum
\begin{equation}
    \label{eq:recoil}
    \mathcal R_k(\delta m_1^2) = \int dE_r\,G_k(E_r)\,\frac{dR}{dE_r},
\end{equation}
where $k$ labels the reconstructed-energy bin. $G_k(E_r)$ is the bin-integrated Gaussian smearing function that maps a true recoil energy $E_r$ into reconstructed-energy bin $k$. In this study, we center the reconstructed energy at $E_{\mathrm {rec}} = E_r$ and model the energy resolution with a Gaussian smearing of width
\begin{equation}
    \sigma_E(E_r)=0.310\sqrt{E_r}+0.0037E_r,
\end{equation}
with $E_r$ and $\sigma_E$ expressed in keV~\cite{XENON:2020rca}. This Gaussian smearing is not intended to reproduce the full XENONnT detector-response model, but instead provides an approximate mapping between the true recoil spectrum and reconstructed energy $E_{\mathrm{ rec}}$~\cite{XENON:2024mlv,XENON:2026rbz}. 

The SM reference is obtained from Eq.~\eqref{eq:recoil} by setting $\delta m_1^2=0$, such that $\mathcal R_k^{\rm SM}\equiv\mathcal R_k^{\rm PD}(\delta m_1^2=0)$. The bin-by-bin deformation is then
\begin{equation}
    \label{eq:rho_def}
    \rho_k(\delta m_1^2) \equiv \frac{\mathcal R_k^{\rm PD}(\delta m_1^2)}{\mathcal R_k^{\rm SM}}.
\end{equation}
By construction, $\rho_k(0)=1$. Any common normalization factors in the theoretical recoil calculation cancel in Eq.~\eqref{eq:rho_def}, while the overall active-rate suppression and the energy-dependent spectral distortion induced by the pseudo-Dirac splitting are retained. We do not normalize $\rho_k$ to unit area, since the reduction of the active solar-neutrino rate is itself part of the pseudo-Dirac signal. For the mass-squared splittings relevant to our constraint, the pseudo-Dirac oscillation produces both an overall suppression and an energy-dependent deformation that is strongest toward the lower-energy portion of the XENONnT recoil spectrum (see the bottom of Fig.~\ref{fig:pd_deformation}). Thus, our inference on $\delta m_1^2$ derives from the overall suppression of the active solar-neutrino rate in addition to the accompanying spectral deformation.

We construct the detector-level XENONnT SM solar-neutrino template from the published solar-neutrino spectral shape in Ref.~\cite{XENON:2026rbz}. For each science run, SR0 and SR1b, labeled by $r$, we digitize the published best-fit solar-neutrino spectrum and denote its normalized spectral shape by $c^\odot_{rk}$, with $\sum_k c^\odot_{rk}=1$. The nominal SM normalization is fixed to the XENONnT nominal solar-neutrino yields $N^{\rm nom}_{s,\mathrm{SR0}}=295$ and $N^{\rm nom}_{s,\mathrm{SR1b}}=387$~\cite{XENON:2026rbz},
\begin{equation}
    \label{eq:solar_template}
    S_{rk}^{\rm SM} = N^{\rm nom}_{\odot,r}c^\odot_{rk}.
\end{equation}
We obtain the pseudo-Dirac signal template by applying the theoretical deformation bin by bin to the published XENONnT solar-neutrino template, neglecting the sub-percent contribution from the remaining solar components,
\begin{equation}
    \label{eq:signal_PD}
    S_{rk}^{\rm PD}(\delta m_1^2)= S_{rk}^{\rm SM}\rho_k(\delta m_1^2).
\end{equation}

\subsection{Likelihood Analysis}

\begin{figure}[t]
    \centering
    \includegraphics[width=0.9\linewidth]{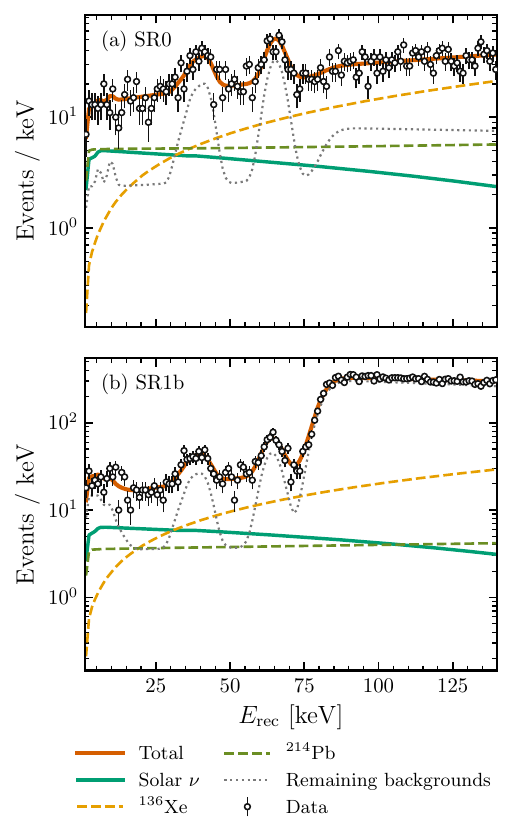}
    \caption{Reproduction of the published XENONnT SR0 spectrum (a) and SR1b spectrum (b). The digitized published solar-neutrino spectra integrate to 501 events in SR0 and 658 events in SR1b, in close agreement with the published best-fit yields of 500 and 660 events, respectively~\cite{XENON:2026rbz}.}
    \label{fig:xenon_reconstruction}
\end{figure}

The official XENONnT analysis employs a joint unbinned extended likelihood in reconstructed energy~\cite{XENON:2026rbz}. Although XENONnT publishes the form of this likelihood, the event-level data and full numerical likelihood ingredients are not publicly available. We therefore construct a binned approximation from the published SR0 and SR1b spectra. We use 139 one-keV bins over the reconstructed-energy range $1\,\mathrm{keV}\leq E_{\mathrm{rec}}<140\,\mathrm{keV}$. Let $n_{rk}$ denote the observed number of events in run $r$ and reconstructed-energy bin $k$, obtained by digitizing the published XENONnT science-data points (see Fig.~\ref{fig:xenon_reconstruction}). We likewise digitize the published best-fit background-component spectra to construct the normalized background spectral shapes. Our recast retains separate detector-level shapes for ${}^{214}$Pb, ${}^{212}$Pb, ${}^{85}$Kr, material backgrounds, ${}^{136}$Xe, and the SR1b ${}^{3}$H-like component from the XENONnT publication. The remaining subdominant backgrounds displayed collectively by XENONnT as ``Other background'' are represented by the corresponding published grouped spectrum over the full analysis range. The nominal XENONnT background model does not include a ${}^14$C component. We note that XENONnT separately considers ${}^{14}$C contamination as a conservative alternative-background scenario, whose inclusion could weaken the inferred pseudo-Dirac constraint~\cite{XENON:2026rbz}.

The expected number of events in run $r$ and reconstructed-energy bin $k$ is
\begin{equation}
    \label{eq:lambda}
    \lambda_{rk} = S_{rk} + \sum_b \mu_{b,r}c^b_{rk},
\end{equation}
where $b$ runs over the background components. Here $c^b_{rk}$ is the normalized spectral shape of background component $b$, $\mu_{b,r}$ is its expected yield in run $r$, and $S_{rk}$ is the solar-neutrino signal contribution. For the observed counts $n_{rk}$, we use a binned Poisson likelihood. Background nuisance parameters with published constraints are assigned Gaussian constraint terms centered on their nominal expectations~\cite{XENON:2026rbz}. We write the likelihood schematically as
\begin{equation}
    \label{eq:binned_likelihood}
    \mathcal L = \prod_{r,k}\frac{e^{-\lambda_{rk}}\lambda_{rk}^{n_{rk}}}{n_{rk}!}\prod_{m}\exp\left[-\frac{(\theta_m-\bar{\theta}_m)^2}{2\sigma_m^2}\right],
\end{equation}
where $\theta_m$ denotes a constrained nuisance parameter, with nominal value $\bar{\theta}_m$ and uncertainty $\sigma_m$. The ${}^{214}$Pb, ${}^{212}$Pb, ${}^{85}$Kr, and material-background yields are constrained independently in each run according to their published uncertainties. The ${}^{136}$Xe component is assigned a common fractional normalization across SR0 and SR1b, while the SR1b ${}^{3}$H-like component and the grouped remaining backgrounds are allowed to float freely. All nuisance parameters are profiled simultaneously in each fit.

To validate our public-data recast, we temporarily set $\rho_k=1$ and replace the solar-neutrino signal normalization by a free common signal-strength parameter $\mu_\odot$, such that
\begin{equation}
    S_{rk}=\mu_\odot S_{rk}^{\rm SM}.
\end{equation}
After profiling over the background nuisance parameters, the likelihood is maximized at $\hat{\mu}_\odot=1.76$, consistent with the upward solar-neutrino signal-strength preference reported by XENONnT. The relative profile-likelihood width is $13.8\%$, reproducing the scale of the XENONnT solar-neutrino profile-likelihood uncertainty~\cite{XENON:2026rbz}. The free parameter $\mu_\odot$ is used only for this validation and is not further retained.

For the pseudo-Dirac fit, we normalize the solar-neutrino contribution to the XENONnT nominal SM yield. XENONnT additionally reports a $14\%$ uncertainty from non-rate systematic effects~\cite{XENON:2026rbz}. As the full detector systematic model is not publicly reconstructible, we approximate this effect with a constrained solar signal-normalization parameter,
\begin{equation}
    S_{rk} = \eta_\odot S_{rk}^{\rm PD}(\delta m_1^2),
\end{equation}
where $\eta_\odot$ is a constrained solar signal-normalization nuisance parameter with $\eta_\odot=1\pm0.14$, implemented through a Gaussian constraint term.

\begin{figure}[ht]
    \centering
    \includegraphics[width=0.9\linewidth]{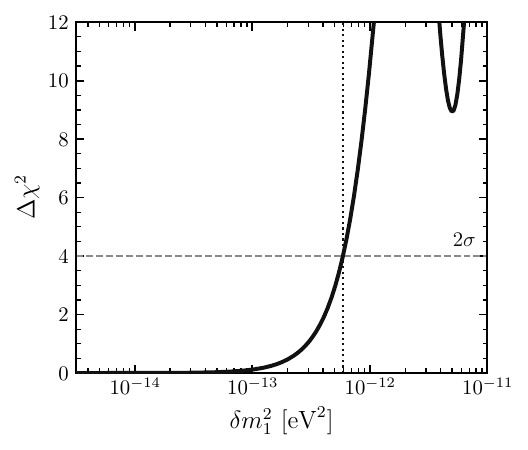}
    \caption{Adopting $\Delta\chi^2=4$ as the $2 \sigma$ criterion, the $\Delta \chi^2$ profile first crosses at $\delta m_1^2=5.90\times10^{-13}\,\mathrm{eV}^2$ and remains above it through the upper endpoint $\delta m_1^2 = 10^{-11}\,\mathrm{eV}^2$, with no secondary allowed region.}
    \label{fig:dm1_profile}
\end{figure}

At each fixed value of $\delta m_1^2$, we profile over all nuisance parameters,
\begin{equation}
    \mathcal L_{\rm prof}(\delta m_1^2) = \max_{\{\theta_m\}} \mathcal L(\delta m_1^2,\{\theta_m\}),
\end{equation}
and define the likelihood-ratio test statistic~\cite{Cowan:2010js}
\begin{equation}
    \Delta\chi^2(\delta m_1^2) = -2\ln\left[ \frac{\mathcal L_{\rm prof}(\delta m_1^2)} {\mathcal L_{\max}} \right],
\end{equation}
where $\mathcal L_{\max}$ is the global maximum of the profiled likelihood over the scanned values of $\delta m_1^2 \leq 10^{-11}\,\mathrm{eV}^2$. The global best fit occurs at $\delta m_1^2=0$. Adopting $\Delta\chi^2=4$ as the one-parameter $2\sigma$ criterion, we exclude $5.90\times10^{-13} \,\mathrm{eV}^2 < \delta m_1^2 \leq 10^{-11}\,\mathrm{eV}^2$ in the interval considered, as shown in Fig.~\ref{fig:dm1_profile}. 

\section{Discussion}
The XENONnT $pp$-dominated spectrum demonstrates the potential for current and future multiton dark-matter experiments to also collect statistically useful samples of low-energy solar neutrinos~\cite{DARWIN:2020bnc, JUNO:2023zty,PandaX:2026qdp,XENON:2026rbz}. As dark-matter experiments continue to improve their low-energy exposures and control of backgrounds, their sensitivity to the solar-neutrino flux opens an additional avenue for precision neutrino physics. Using the published XENONnT SR0 and SR1b spectra, we find that the best fit occurs at zero pseudo-Dirac splitting and exclude $5.9\times10^{-13}\,\mathrm{eV}^2 \lesssim \delta m_1^2 \leq 10^{-11}\,\mathrm{eV}^2$ at $2 \sigma$. Thus, the XENONnT public data extend the lower edge of the solar-neutrino constraint on $\delta m_1^2$~\cite{Ansarifard:2022kvy}, bringing current solar-neutrino sensitivity toward the mass-splitting regime previously emphasized in forecasts for future $pp$-sensitive experiments~\cite{deGouvea:2021ymm,Franklin:2023diy}. Additional low-energy backgrounds, including the conservative ${}^{14}$C scenario considered separately by XENONnT, could weaken the stated constraint. This motivates dedicated pseudo-Dirac sensitivity studies for multiton dark-matter detectors, where larger solar-neutrino samples and improved control of low-energy backgrounds could extend the reach to even smaller mass-squared splittings. 

The present excluded interval is obtained from the published energy spectra and integrated run exposures. The upper endpoint of this interval defines the conservative domain of the stated public-data recast rather than a fundamental upper boundary of pseudo-Dirac phenomenology. Hence, a public time-resolved XENONnT solar-neutrino data set would make it possible to include the annual variation of the Earth--Sun baseline and test the corresponding phase-dependent modulation directly~\cite{BOREXINO:2022wuy,Ansarifard:2022kvy}. A complementary analysis incorporating higher-energy ${}^8$B solar neutrinos could probe a pseudo-Dirac splitting associated with the second standard mass eigenstate~\cite{Ansarifard:2022kvy,XENON:2026smt}. Finally, relaxing the maximal active--sterile mixing of the pseudo-Dirac limit would extend the analysis to a more general ultra-light $1$--$4$ sterile-neutrino scenario~\cite{Chen:2022zts,deGouvea:2021ymm,Goldhagen:2021kxe}. We leave these extensions for future work.

\begin{acknowledgments}
We thank Rocky Kolb and Gordan Krnjaic for thoughtful feedback. GL was supported by a generous fellowship from the University of Chicago Department of Physics for the duration of this work.
\end{acknowledgments}

\bibliography{biblio.bib}

\end{document}